# AN ECONOPHYSICS MODEL FOR THE STOCK-MARKETS' ANALYSIS AND DIAGNOSIS

Ion SPÂNULESCU[*], Ion POPESCU[**],
Victor STOICA[*], Anca GHEORGHIU[*] and Victor VELTER[***]

*Abstract.* In this paper we present an econophysic model for the description of shares transactions in a capital market. For introducing the fundamentals of this model we used an analogy between the electrical field produced by a system of charges and the overall of economic and financial information of the shares transactions from the stock-markets. An energetic approach of the rate variation for the shares traded on the financial markets was proposed and studied.

## 1. Introduction

The stock exchange rate of a share or other quoted title is the price at which those titles exchange on stock markets. This rate varies according to the law of supply and demand, and the rate approaches more or less the true value of the share.

The setting of the stock exchange rate is based on all available information about the asset proposed for trading, primarily on the intrinsic value estimated by different methods of evaluation of the company or economic entity which is traded on the stock-market. These evaluation methods are based on certain expectations, therefore contain a certain amount of subjectivity depending on the evaluators' quality, and the model used for it [4]. Therefore, these methods can only give an estimated potential rate, also called the **intrinsic value**, which could possibly help the stock investors in making buying or selling decisions. The problem of the real value arises when the absence of quotation of the asset to the stock market is higher, i.e. the reference price is not fixed yet; in this case

---

[*] Hyperion University of Bucharest, 169 Calea Călăraşilor, St., Bucharest, 030615.
[**] Spiru Haret University, 13 Ion Ghica, St., Bucharest, Romania.
[***] Valachia University, 2 Carol I, St., Târgovişte, 130024, Dâmboviţa County, Romania.

theoretical estimates may serve as a basis for negotiation when taking into account for trading.

In this article, we propose a theoretical model to estimate rate of the share and its evolution over time depending on supply and demand. This model is based on the similarity between an electrostatic field – symbollizing the stock-market (or field) and the electrical charges which symbollizes all the assemble of the economic information about the transactioned shares.

## 2. The Stock-Markets' Electrostatic Model

The market value of a share at a certain time $t$ is an average, steady value, as determined in the negotiated market transactions. The principle followed is that of ensuring counterpart; according to this principle the purchase orders sent to a higher price and the sell orders submitted at a lower price than the market price have their chances of implementing. The accumulation/aggregation of the purchase orders is made from the highest to the lowest rate; the aggregation of the sales orders is made in reverse; the buying and selling orders to the market are primarily/mainly taken into account and are executed at the market rate which is the balance rate (value) at a certain time, when the number is comparable to the offered one.

The aggregate evolution of the securities rates on a certain market means the range of values taken by the stock market-rates aggregated in a certain time period subjected to observation on daily basis or in a one day stock-market session (Fig. 1,*c*).

Let's consider the shares $X_i$ of a company $C$ listed on any stock exchange market.

Suppose that on day $z + 1$, the stock rate of the shares $X_i$ increases from 950 monetary units (m.u.) to 1,000 m.u. This growth rate is determined by the establishment of equilibrium between the supply and request/demand at a higher level, thus moving the rate from 950 m.u. on the day $z$ to 1,000 m.u. on the day $z + 1$. On the day $z + 2$, the rate goes up again to a value of 1,100 m.u. and so on.

We consider the range of scalar values of equilibrium of the rate of share $X_i$ as similar to an electrostatic field $E_i$ produced by the charge $+ q_i$ symbolizing – in economic terms – all the economic information (price, volume, type of share and so on) about the share $X_i$ (Fig. 1,*a*,*b*).

This can be generalized considering that all information about all shares traded in the stock market A, are sources $a_i$ for an information field $E_b$ symbolizing the whole capital market (Fig. 1,c).

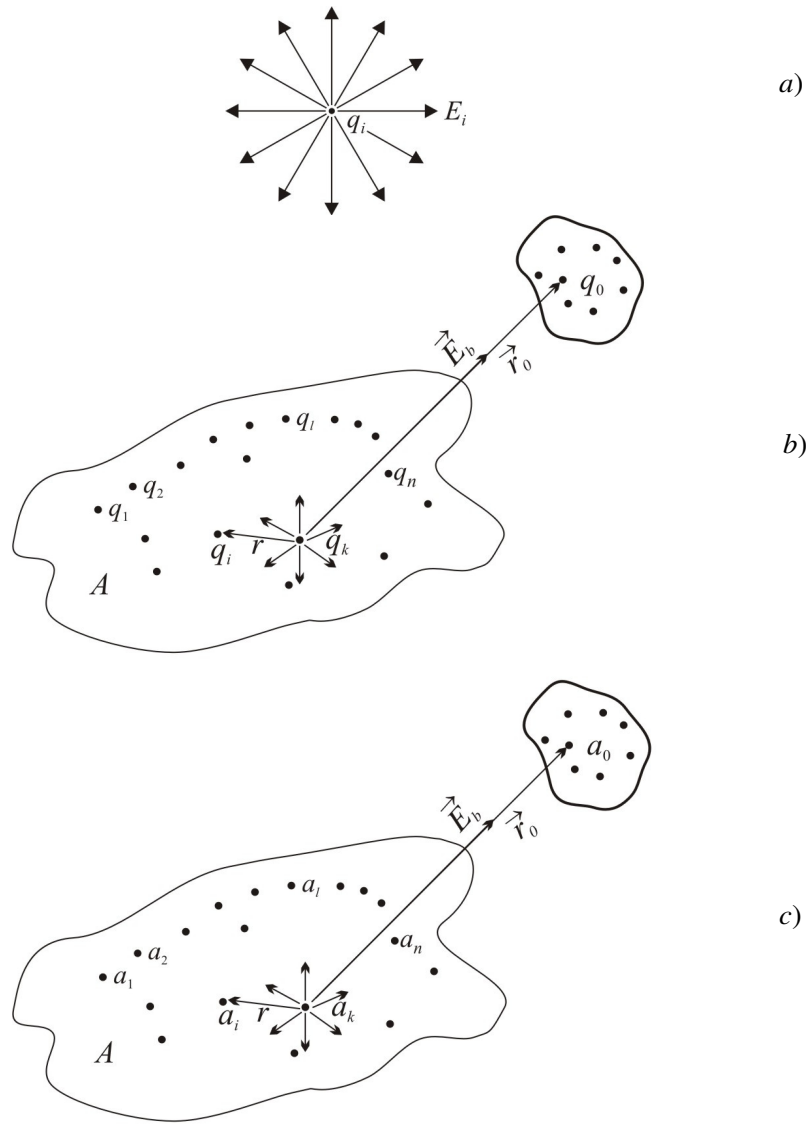

**Figure 1.** a) Electrostatic field created by electrical change $+q_i$; b) Electrostatic field created by electrical charges system $q_1, q_2, …q_n$; c) The stock-market information field created by $a_1, a_2 …, a_n$ information sources.

As the electrostatic field, the stock-market information field is a force field whose value is proportional to field intensity, $E$. For example, in physics the electrical force $F_e$ is [1]:

$$F_e = q_0 E \tag{1}$$

where $q_0$ represents an electrical charge. Similarly, the force of the stock-market quotation of shares, $F_b$, is given by the law of supply and request, and can be expressed similar to (1):

$$F_b = R = aE_b \qquad (2)$$

where $a$ represents „the charge" of the point symbolizing the **equilibrium value** of a share $X_i$ in the information field.

As in physics, to keep a material particle or electrical charge in a fixed point, the electrical force $F_e = qE$ is compensated by a mechanical force $F_m$, in the stock-market field acts the law of supply and demand, similar to the law of action and reaction in physics (Fig. 2,*a*).

The electrostatic field created by the charges $q_1$, $q_2$, ...$q_n$, interacts with the charge $q_0$ located at the distance $r_0$ (Fig. 1,*b*) through the Coulomb type electrical force (1) where $F_e$ is given by Coulomb's Law [1]:

$$F_e = \frac{1}{4\pi\varepsilon_0} \frac{q_0 \sum_{j=1}^{n} q_j}{r_0^2}. \qquad (3)$$

The total charge $Q$ inside the region $A$ is given by the sum $\sum_{j=1}^{n} q_j$, and $1/4\pi\varepsilon_0$ represents the proportionality constant expressed in the International System of units used in electricity [1]. Comparing the equations (1) and (3), we obtain the equation for the electrical (or electrostatic) field:

$$E = \sum_{j=1}^{n} \frac{1}{4\pi\varepsilon_0} \frac{q_j}{r_0^2} \qquad (4)$$

or:

$$E = k \frac{Q}{r_0^2}. \qquad (5)$$

In Eq. (5) $Q$ represents the total charge from region $A$ and $r_0$ is the distance between the region $A$ which creates the field $E$ and the point $P_0$ where is located the charge $q_0$ that the field interacts with (Fig. 1,*b*). We notice that the fields' intensity, thus the influence over $q_0$ charge is smaller as the distance between the point $P_0$ and the region $A$ is larger (Fig. 1,*c*).

Similarly, with Eq. (5) we can write equation for the stock-market information field $E_b$ (Fig. 1,*c*):

$$E_b = k_b \frac{Q_b}{r_0^2} \qquad (6)$$

where $r_0$ represents the so called information, „distance" between the field produced by the figurative points that represent the information „charges" for the shares contained by a market $A$ (Fig. 1,*c*) and other share on another market located at the informational „distance" $r_0$ (Fig. 1,*c*). It is obvious that this „distance" will be as longer as big are the difference between the economic and financial information about „the shares", like the difference between the energy or oil shares and the financial shares for banks or investment funds and so on [2].

Analyzing the region $A$ from figure 1,*c*, we can see that the information distances $r$ between the same type of shares (financial or banking, for example) are much shorter, so their interaction with the stock-market field is much stronger and the fields' intensity is much higher:

$$E_b = k_b \frac{Q_b}{r^2} \qquad (7)$$

because, as seen in figure 1,*b*,*c*, we have $r \ll r_0$. In this latter case, considering the fields' interaction $E_b$ with the $q_i$ charge from within the (stock-market) region $A$, the force of the share given by the equation:

$$F_k = k_b \frac{q_i Q}{r^2} \qquad (8)$$

is much stronger because of much smaller information „distances" $r$ comparing with distance $r_0$.

## 3. The Energetic Approach of the Stock-Market Rate

As is known in physics, the fields' forces can produce mechanical work when acting on a material point or electrical charge $q$ on a finite distance $\Delta S$ etc.:

$$L = F \Delta S. \qquad (9)$$

The electrostatic field $E$ created by the $+q$ charge, will interact with a test-charge $+q_0$ with a rejection force $F_e = q_0 E$ seeking to push away the positive $+q_0$ charge. For the $+q_0$ charge to remain motionless (in electrostatic terms, standing) is necessary to action with a mechanical force $F_m$ equal and opposite with the electrical force $F_e$ of the electric field i.e. (Fig. 2,*a*):

$$\vec{F}_m = -\vec{F}_e = -q_0 \vec{E}. \qquad (10)$$

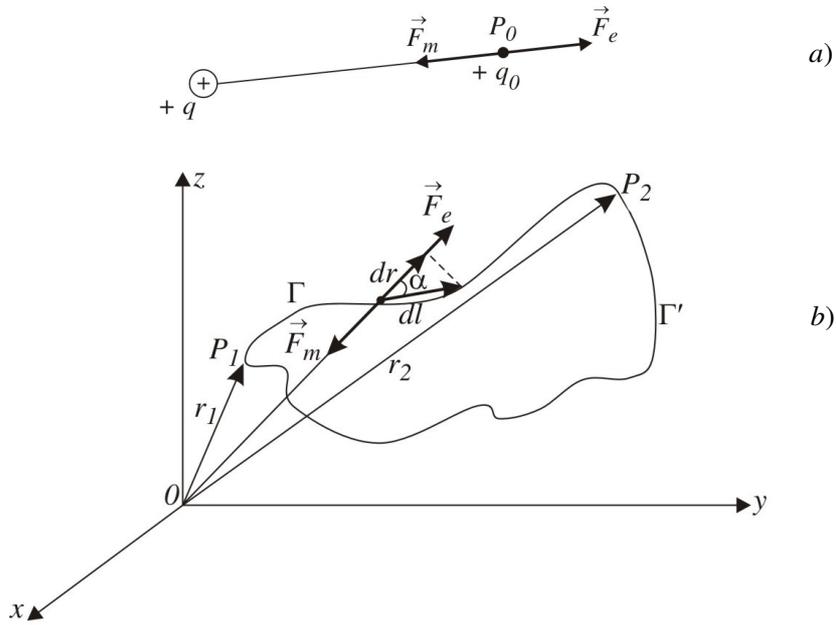

**Figure 2.** a) Electrical rejection force $F_e$ is compensate by the mechanical force $F_m$; b) Calculation of mechanical work of force $F_e$ betwen $P_1$ and $P_2$ points.

The elementary mechanical work d$L$ done by the force $F_m$ when the charge $q_0$ is moving on the d$l$ displacement (see Fig. 2,b) is given by [1]:

$$dL = \vec{F}_m(\vec{r})d\vec{l} = -q_0\vec{E}(\vec{r})\,d\vec{l}, \tag{11}$$

where $\vec{F}_m(\vec{r})$, $d\vec{l}$ and $\vec{E}(\vec{r})$ are vector parameters.

The mechanical work of the force $F_m$ between the points $P_1$ and $P_2$ (Fig. 2,b) is given by:

$$L_{1,2} = \int_1^2 dL = \int_{r_1}^{r_2} \vec{F}_m(\vec{r})d\vec{l} = -q_0 \int_{r_1}^{r_2} \vec{E}(\vec{r})d\vec{l}. \tag{12}$$

Taking the axes origin in the center of the source-charge $+q$ and taking into account the expression (4) will find (following the calculations see [1], p. 39), the equation for the mechanical work:

$$L_{1,2} = -q_0 \frac{q}{4\pi\varepsilon_0} \int_{r_1}^{r_2} \frac{dr}{r^2} = \frac{qq_0}{4\pi\varepsilon_0}\left(\frac{1}{r_2} - \frac{1}{r_1}\right) \tag{13}$$

or:

$$L_{1,2} = kq_0q\left(\frac{1}{r_2} - \frac{1}{r_1}\right). \tag{14}$$

The equation (14) shows that the mechanical work $L_{1,2}$ done for the displacement of the constant charge $q_0$ in the field created by the $q$ charge from origin **does not depend on the path** followed between the points $P_1$ and $P_2$, but only on the initial and final position, it means it depends only on the initial distance $r_1$ and the final distance $r_2$ between origin of axes and the point $P_1$ and $P_2$ respectively (Fig. 2,*b*). It is known from mechanics that such a field having this property (that the mechanical work depends only on initial and final points) is a **conservative field**, that means it possesses potential energy on whose behalf the mechanical work is carried out by the field forces.

Similarly it can be considered that the information stock-market field $E_b$ is a conservative field where operates the law of supply and request in the stock-market. The work done by the stock-market field forces is given by an equation similar with Eq. (9):

$$L_b = R\Delta P \qquad (15)$$

being determined by the stock-market field forces $R$, i.e. by the action of the law of request (Request = $R$) and offer and by the rate variation $\Delta P$ of the share price.

In equation (15), $R$ represents the stock market force $F_b$ given by the equation (see Eq. (2)):

$$F_b = R = aE_b \qquad (16)$$

where $E_b$ is the information field of the equilibrium values and other information regarding the shares, and $a$ represents the „charge" of the point symbolizing the all information and the shares' value $X_i$. The size of $R$ (Request) represents the reliability degree or appreciation of the considered share that also determines the level of request $R$ of the share.

Finally, under the action of the field $E_b$, of the force $R$, by raising the request and the offer, and hence the share rate (value), a stock-market „work" is performed given by the equation (see Eq. (9)):

$$L_b = R\Delta P = aE_b\Delta P \qquad (17)$$

where $\Delta P$ represents the share's rate variation.

For the **conservative** fields, like the electrostatic field, the mechanical work performed at the movement of the charge $q_0$ from the point $P_1$ to the point $P_2$ (Fig. 2,*b*) is given by the difference between the potential energies of the charge $q_0$ in the two points, and that is [1]:

$$L_{1,2} = W(r_2) - W(r_1) \qquad (18)$$

where $W(r_2)$ and $W(r_1)$ are the potential energies of the „test" charge in the points $P_1$ and $P_2$ (Fig. 2,b). Similarly, in the stock-market a virtual point symbolizing all the information about the share $X_i$ from a forces potential field $E_b$, like the capital market, possesses the stock-market **potential energy** reflected in its price or stock-market rate increase **opportunities** on whose behalf the stock-market "mechanical" work is effectuated (Fig. 3):

$$L_{b(1,2)} = W(v_2) - W(v_1). \qquad (19)$$

If there is a variation of the share rate, from $v_1 = 950$ m.u. to the higher equilibrium value, for example $v_2 = 1000$ m.u., in this case the equation (19) for the stock-market work is:

$$L_{b(1,2)} = W(1000) - W(950).$$

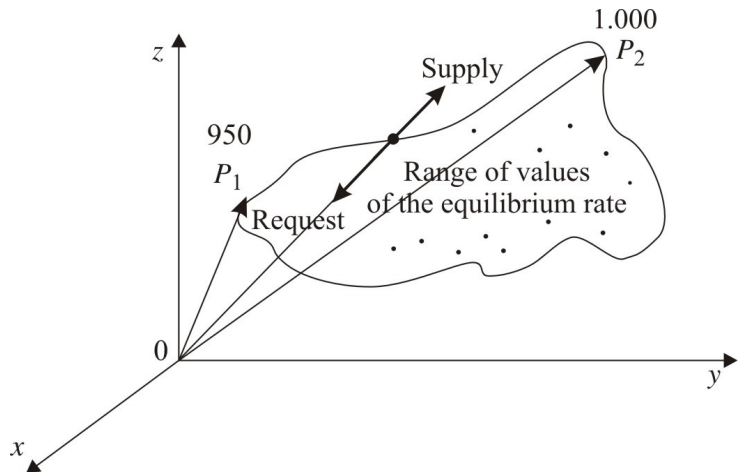

**Figure 3.**

As in mechanics or electrostatics, if the stock-market rate is decreesing, there is a transformation of the potential energy in kinetic energy, and vice versa for the increasing of the rate. The maximum kinetic energy is achieved in the crush moment until reaching the minimum resistance, than the process can be resumed depending on market development of the value of the share $X_i$ (Fig. 4).

As it can be seen in figure 4, the successive variations of the transformation of the potential energy into stock-market kinetic energy can take place several times in a one day stock-market session or in other time period represented by the graphics of the evolution of share values. The minimum value of the potential energy $W_{pb}$ is when the shares' rate value is minimum, thus the share $X_i$ has in that moment a low "potential". In exchange, in that moment, the kinetic stock-market energy $E_{cb}$, depending to the variation (raising) **speed** of the rate and to the transaction time, is high, the shares' value can change rapidly, i.e. to increase (Fig. 4) or to begin to decrease again until it reaches a new resistance threshold of the rate.

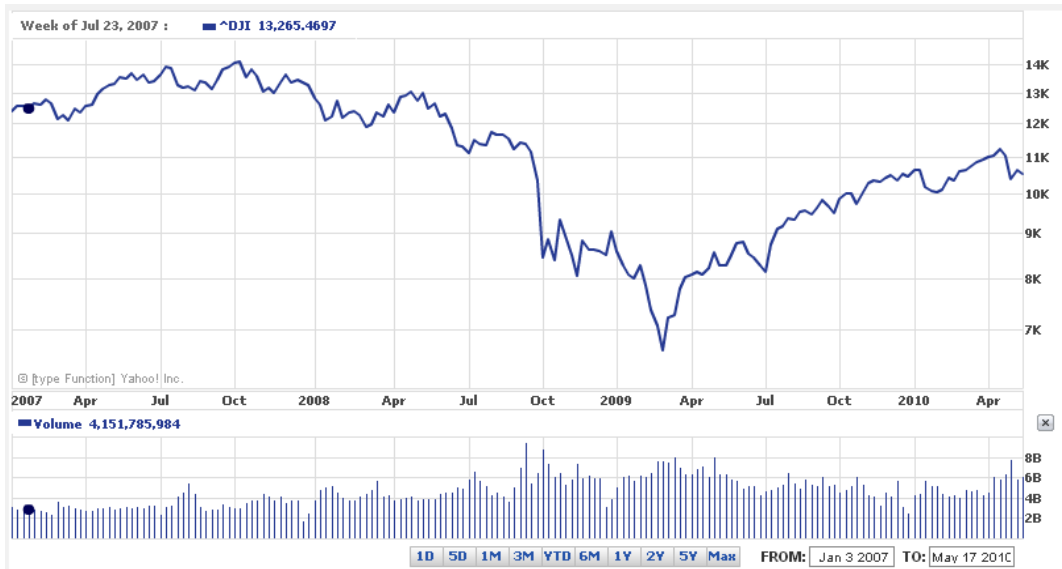

**Figure 4.** Weekly evolution of DJI.

## 4. Conclusions

By similarity with the electrostatic field notion, we can introduce an information field concept for modeling the stock-market shares transactions.

The stock-market information field sources are the points symbolizing the totality of economic and financial information about the shares listed at that time.

Similarly with the electrical force notion given by the equation $F_e = qE$, where $E$ is an electric field, we propose the introduction of the notion of stock-market force equal with the request $R$ or with the supply, and being proportional with the stock-market fields' intensity with an information nature, i.e. $R = aE_b$.

The "mechanical" work executed by the forces of the stock-market field $L_b$ is given by the product between the *R force, determined by the law of supply and request and by the price variation* $\Delta P$ (the share's rate variation), i.e.: $L_b = R\Delta P$.